\documentclass[apj]{emulateapj}
\usepackage{graphicx}
\usepackage{epstopdf}
\usepackage{amsmath}
\usepackage{amssymb}
\shorttitle{AGN feedback}
\shortauthors{Bu \& Yang}
\begin{document}
\title{Active galactic nuclei feedback at parsec scale}
\author{De-Fu Bu\altaffilmark{1}, Xiao-Hong Yang\altaffilmark{2}}

\altaffiltext{1}{Key Laboratory for Research in Galaxies and Cosmology, Shanghai Astronomical Observatory, Chinese Academy of Sciences, 80 Nandan Road, Shanghai 200030, China; dfbu@shao.ac.cn }
\altaffiltext{2}{Department of physics, Chongqing University, Chongqing, 400044; yangxh@cqu.edu.cn}

%\date{Accepted 1988 December 15. Received 1988 December 14; in original form 1988 October 11}

%\pagerange{\pageref{firstpage}--\pageref{lastpage}} \pubyear{2002}

%\maketitle

%\label{firstpage}
%
\begin{abstract}
We perform simulations to study the effects of active galactic nuclei (AGNs) radiation and wind feedback on the properties of slowly rotating accretion flow at parsec scale. We find that when only radiative feedback is considered, outflows can be produced by the radiation pressure due to Thomson scattering. The mass flux of outflow is comparable to that of inflow. Although strong outflow is present, the luminosity of AGN can be easily super-Eddington. When wind feedback is also taken into account, the mass flux of outflow does not change much. Consequently, the luminosity of the central AGN can still be super-Eddington. However, observations show that the luminosity of most AGNs is sub-Eddington. Some other mechanisms are needed to reduce the AGNs luminosity. Although the mass outflow rate does not change much by wind feedback, other properties of outflow (density, temperature, velocity and kinetic power) can be significantly changed by wind feedback. In the presence of wind feedback, the density of outflow becomes significantly lower, temperature of outflow becomes significantly higher, velocity of outflow is increased by one order of magnitude, the kinetic power of outflow is increased by a factor of $40-100$.
\end{abstract}


\section{Introduction}
It is widely believed that a super massive black hole (SMBH) is located at the center of almost each galaxy (see Heckman \& Best 2014 for a review). The SMBH is believed to co-evolve with its host galaxy (Hopkins et al. 2005; Cattaneo et al. 2009). There are many observations showing that the mass of the SMBH is tightly correlated with the properties of its host galaxy (Fabian 2012). One of the example is that the mass of the SMBH is a constant fraction of the stellar bulge mass (H$\ddot{\rm a}$ring \& Rix 2004). The even more remarkable evidence is that the mass of the SMBH is tightly correlated with the velocity dispersion of the host galaxy's central bulge (e.g., Magorrian et al. 1998; Ferrarese \& Merritt 2000; Gebhardt et al. 2000; Tremaine et al. 2002; G$\ddot{\rm u}$ltekin et al. 2009; Kormendy \& Ho 2013). AGNs feedback plays an important role in the evolution of its host galaxy (Fabian 2012; king \& Pounds 2015).

There are three kinds of output by an AGN, namely, radiation, wind and jet. The radiation from an AGN can Compton heat/cool the interstellar medium (Ciotti \& Ostriker 1997, 2001, 2007; Ciotti et al. 2009). It is found that when the accretion flow is in hot mode, the gas in the region around the Bondi radius can be heated up to be above the virial temperature. Consequently, winds can be generated due to Compton heating around Bondi radius (Bu \& Yang 2018, 2019a; Yang \& Bu 2018). For black hole X-ray binaries (BHBs), it is also found that wind can be driven from accretion disk at large radii due to the Compton heating (Higginbottom et al. 2018, 2019). The line force due to the absorption of UV photons by the low ionization gas can also launch strong wind (e.g., Proga et al. 2000; Murray et al. 1995; Liu et al. 2013; Nomura et al. 2016). AGN winds have large opening angle and can interact with the gas surrounding the AGN very efficiently. Gas surrounding an AGN can be directly blown away by winds, which will affect the star formation rate and the AGN fueling (e.g., Ostriker et al. 2010; Du et al. 2014; Weinberger et al. 2017, 2018; Ciotti et al. 2017; Yuan et al. 2018; Bu \& Yang 2019b). Due to the very small opening angle of jet, it is believed that jet can not effectively interact with gas in the region around an AGN. Jet should play an important role in galaxy cluster scale (Guo 2016; Guo et al. 2018).

Kollmeier et al. (2016) studied the luminosity distribution of a sample of 407 AGNs in the redshift range $z \sim 0.3-4$. It is found that the distribution of the estimated Eddington ratios can be well described as log-normal, with a peak at $L_{\rm bol}/L_{\rm Edd} \approx 1/4$ ($L_{\rm bol}$ and $L_{\rm Edd}$ being bolometric and Eddington luminosity, respectively). In other words, almost all the AGNs in this smaple have sub-Eddington luminosity. Later, Steinhardt \& Elvis (2010) used a even larger sample composed of 62185 quasars from the Sloan Digital Sky Survey Data to study the luminosity distributions of quasars. They made the same conclusion as that in Kollmeier et al. (2016). The fuelling gas at galactic scale has a very broad mass distribution (Liu et al. 2013). In other words, the galactic scale inflow can provide sufficient gas to make the AGN's luminosity super-Eddington. Also, the black hole slim accretion disk model predicts that the rotating accretion disk can have well higher luminosity above $L_{\rm Edd}$ (Abramowicz et al. 1988; Ohsuga et al. 2005; Yang et al. 2014). What is the reason for the sub-Eddington accretion of the AGNs?

AGN radiative and wind feedback may be possible ways to solve the `sub-Eddington puzzle'. In the innermost region very close the central black hole, the numerical simulations of slim disk (Ohsuga et al. 2005) with radiative transfer show that the accretion disk can radiate well above $L_{\rm Edd}$. Therefore, the radiative feedback in the innermost region can not be effective in reducing the mass accretion rate (or equivalently luminosity). On the parsec scale, the effects of AGN radiative feedback on the accretion flow have been studied (kurosawa \& Proga 2009; Liu et al. 2013). It is shown that even though the line force can drive strong wind, the accretion rate (or luminosity) of AGN is still super-Eddington. The sub-Eddington puzzle can not be solved by radiative feedback.

The motivations of the present paper are twofold. The first one is to study whether AGN wind feedback at parsec scale can solve the sub-Eddington puzzle. The second motivation is to study the effects of wind feedback on the properties of the accretion flow at parsec scale. The accretion flow at parsec scale connects galactic inflow and the accretion flow onto the central black hole. It is vitally important to study the accretion flow in this region, because it is in this region that the inflow rate to the central black hole is determined.

In Bu \& Yang (2019b), we studied the effects of feedback of wind from low-luminosity AGN (LLAGN) on properties of accretion flow at parsec scale. In the present paper, we study the high accretion rate luminous AGN case by setting much higher gas density at the outer boundary. The properties of wind from LLAGN are quite different from those of wind from luminous AGNs (see the details in Sections 2.2 and 2.3). Therefore, we expect that the effects of wind feedback by LLAGN should be quite different from those of wind feedback by luminous AGN.

The structure of this paper is as follows. In section 2, we present the numerical settings of simulations and the physical assumptions. The results are presented in section 3; We summarize our results in Section 4.

\section{Numerical method }
The mass of the central black hole is set to be $M_{\rm BH}=10^8M_{\odot}$ with $M_{\odot}$ being solar mass. We study the low angular momentum accretion flow in the region $1500r_s\leq r \leq 7.5 \times 10^5r_s$ ($r_s$ is Schwarzschild radius). In this paper, we only calculate the region above the midplane, we have $0 \leq \theta \leq \pi/2$. We set the specific angular momentum of gas to equal to the Keplerian angular momentum at $r_c=900r_{\text{s}}$, which is smaller than our inner radial boundary.

By using the ZEUS-MP code (Hayes et al. 2006), we run two-dimensional hydrodynamic simulations in spherical coordinates ($r,\theta,\phi$). The equations describing the accretion flow are as follows,

\begin{equation}
 \frac{d\rho}{dt} + \rho \nabla \cdot {\bf v} = 0,
\end{equation}
\begin{equation}
 \rho \frac{d{\bf v}}{dt} = -\nabla p - \rho \nabla \Phi + \rho \bf{F_{rad}}
\end{equation}
\begin{equation}
 \rho \frac{d(e/\rho)}{dt} = -p\nabla \cdot {\bf v} + \dot E
\end{equation}
where, $\rho$ is density, $\bf v$ is velocity and $e$ is internal energy. Ideal gas equation is adopted $p=(\gamma-1)e$ with $\gamma=5/3$. $\Phi$ is the gravitational potential of the central black hole. $\bf{F_{rad}}$ is the radiation pressure due to Thomson scattering. $\dot E$ is the net heating rate per unit volume.

For the radiation force term $\bf{F_{rad}}$, we only consider the radiation pressure due to Thomson scattering. The central AGN emits most of its energy in the region very close to the black hole. The inner radial boundary of our simulations is significantly larger than the AGN emitting region, therefore, we can treat the AGN emitting region as a point source.
The radiation force only has the radial component,
\begin{equation}
{F_{\rm rad, r}}=\frac{\kappa_{es}}{c}\left[F_{\rm X}+F_{\rm UV}\right]
\end{equation}
$F_{\rm X}$ and $F_{\rm UV}$ are the X-ray flux and UV flux from the central AGN, respectively. We set $\kappa_{es}=0.4 \ {\rm cm^2 \ g^{-1}}$.

We define the ionization parameter $\xi$ as,
\begin{equation}
\xi=\frac{4\pi F_X}{n}=f_X L_{\rm bol} \exp(-\tau_X)/n r^2
\end{equation}
$f_X$ is the fraction of the total luminosity ($L_{\rm bol}$) in X-ray. $\tau_X=\int_0^r \rho \kappa_{es} dr$ is the X-ray scattering optical depth.

The net heating rate $\dot E$ in Equation (3) includes Compton heating/cooling rate ($S\rm c$), photoionization heating-recombination cooling rate ($G \rm x$), bremsstrahlung cooling rate ($B\rm r$) and line cooling rate ($L_{\rm line}$).
The Compton heating/cooling rate is:
\begin{equation}
S\rm c=8.9\times10^{-36}n^2(T_X-T)\xi \ {\rm erg \ cm^{-3} \ s^{-1}}
\end{equation}
$T_X$ is the Compton temperature of the X-ray photons emitted by the central AGN, $T$ is the temperature of the accreting gas. $n=\rho/(\mu m_p)$ is number density of gas, with $\mu=0.5$ and $m_p$ being mean molecular weight and proton mass, respectively.
The sum of photoionization heating-recombination cooling rate is:
\begin{equation}
G\rm x=1.5\times10^{-21}n^2 \xi^{1/4} T^{-1/2} (1-T/T_X) \ {\rm erg \ cm^{-3} \ s^{-1}}
\end{equation}
The bremsstrahlung cooling rate is:
\begin{equation}
B\rm r=3.3\times 10^{-27}n^2\sqrt{T} \ {\rm erg \ cm^{-3} \ s^{-1}}
\end{equation}
The line cooling rate is:
\begin{equation}
\begin{split}
L_{\rm line}=1.7\times10^{-18}\exp(-1.3\times10^5/T)/\xi/\sqrt{T}+\\
            10^{-24} \ {\rm erg \ cm^{-3} \ s^{-1}}
\end{split}
\end{equation}

In this paper, in all of the models, we take into account the radiative feedback by the central AGN. The radiative feedback includes radiation force $F_{\rm rad, r}$ (Equation (4)), Compton heating/cooling (Equation (6)) and photoionization heating/recombination cooling (Equation (7)). In some models, in addition to radiative feedback, we also take into account wind feedback. The detailed properties of wind are introduced in Sections 2.2 and 2.3.

\subsection{Critical accretion rate}
Observations of BHBs show that when accretion flow luminosity is lower than $2\%L_{\rm Edd}$, the BHBs is in hard state and the accretion flow is hot accretion flow (McClintock \& Remillard 2006). When luminosity is higher than $2\%L_{\rm Edd}$, the BHBs is in soft state and the accretion disk model is standard thin disk (Shakura \& Sunyaev 1973). We use $2\% L_{\rm Edd}$ as the boundary between LLAGN and luminous AGN. Therefore, we defined LLAGN as AGN which has luminosity lower than $2\%L_{\rm Edd}$. The accretion disk model for LLAGN is hot accretion flow (see Yuan \& Narayan 2014 for reviews). Correspondingly, we define luminous AGN as AGN which has luminosity higher than $2\%L_{\rm Edd}$. The accretion disk model for luminous AGN is standard thin disk.

The mass accretion rate corresponding to the critical luminosity is,
\begin{equation}
\dot M_c=\frac{2\%L_{\rm Edd}}{\epsilon_{\rm cold}c^2}
\end{equation}
$\epsilon_{\rm cold}$ is the radiative efficiency of a cold thin disk. In this paper, we set $\epsilon_{\rm cold}=0.1$.

According to the wind model for luminous AGN adopted in this paper, when the central AGN is a luminous AGN, the mass flux of wind is equal to the accretion rate close to the black hole (see Section 2.2.1 for details). In this case, the mass accretion rate at the inner radial boundary of our computational domain ($\dot M_{\rm in}$) is just 2 times the accretion rate at the black hole horizon. Therefore, in order to determine the accretion mode of the central AGN, we directly compare $\dot M_{\rm in}$ with the critical accretion rate $\dot M_c$. When $\dot M_{\rm in}\geq \dot M_c$, the standard thin disk is operating in the central AGN. When $\dot M_{\rm in} < \dot M_c$, the central AGN is a LLAGN and hot accretion flow operates.

\subsection{Cold disk feedback}
\subsubsection{Wind}
Strong winds can be generated from a cold accretion disk. There are many observational evidences showing that strong winds are present in both luminous AGNs (e.g., Crenshaw et al. 2013; Tombesi et al. 2010; Gofford et al. 2015; King \& Pounds 2015; Liu et al. 2014; Liu et al. 2015; Wang et al. 2016; Sun et al. 2017; Park et al. 2018; He et al. 2019) and the soft state of black hole X-ray binaries (e.g., Neilsen \& Homan 2012; Homan et al. 2016; D\'{i}az Trigo \& Boirin 2016; You et al. 2016).

In this paper, we set the mass flux and velocity of wind generated by a cold thin disk according to the observations introduced in Gofford et al. (2015). The reasons for adopting the observational results in Gofford et al. (2015) are as follows. First, the winds are found to be generated at a distance $\sim 100-10^4r_s$ from the black hole (Gofford et al. 2015). The inner boundary of our simulation domain is $1500 r_s$. Therefore, it is reasonable to directly adopt the observational result in our simulations. Second, Gofford et al. (2015) give formulas to describe the properties of wind. Therefore, it is much easier to set the wind properties in simulations.

Gofford et al. (2015) analyzed the properties of wind in a sample composed of 51 Suzaku-observed AGNs. It is found that the wind velocity is a function of the bolometric luminosity of the AGNs,
\begin{equation}
v_{\rm wind}=2.5\times 10^4 \left(\frac{L_{\rm bol}}{10^{45} \ {\rm erg} \ \rm s^{-1}}\right) \ {\rm km} \ {\rm s}^{-1}
\end{equation}
Observations indicate that the wind velocity saturates at a value of $10^5 \ {\rm km} \ \rm s^{-1}$. Accordingly, in our simulations, we set a upper limit of wind velocity at this value.
The mass flux of wind is found to be approximately equal to the accretion rate onto the black hole (Gofford et al. 2015). According to the observation results, we set the mass flux of wind to be equal to the black hole accretion rate. Once we get the mass accretion rate at the inner boundary ($\dot M_{\rm in}$) of the simulation domain, we directly set that the both the wind mass flux and black hole accretion rate as,
\begin{equation}
\dot M_{\rm BH}=\dot M_{\rm wind}=\frac{1}{2}\dot M_{\rm in}
\label{coldmdot}
\end{equation}

The final question to describe wind is its opening angle. In the sample of AGNs analyzed by Gofford et al. (2015), there are some radio-loud AGNs. The properties of the wind in these radio-loud AGNs are also analyzed by Tombesi et al. (2014). Because the inclination angle of jet in these radio-loud AGNs is known, it is easy to know the angle at which wind is present. It is reported in Tombesi et al. (2014) that the winds are detected at a wide range of jet inclination angles ($\sim 10^\circ$ to $70^\circ$, see also Mehdipour \& Costantini 2019). In this paper, we simply inject wind at the inner boundary in the region $10^\circ\leq \theta \leq 70^\circ$. We also assume that the density of wind is uniformly distributed in this region.

\subsubsection{Radiation}
For the luminous AGN ($\dot M_{\rm in}\geq \dot M_c$), we can obtain the black hole accretion rate according to Equation (\ref{coldmdot}). The bolometric luminosity is obtained by using $L_{\rm bol}=\epsilon_{\rm cold} \dot M_{\rm BH} c^2$. We study low-angular momentum gas accretion in this paper. The circularization radius of the gas $r_c$ is smaller than the inner radial boundary of the simulation. We assume that once the gas falls into the inner boundary. A cold thin disk will form at $r_c$. Observations show that a hot corona is present at approximately 10-20 gravitational radius (Reis \& Miller 2013; Uttley et al. 2014; Chainakun et al. 2019), which emits X-ray photons. In this paper, we also assume that a spherical X-ray emitting hot corona is present in the inner most region above and below the thin disk. We assume the X-ray flux is spherically distributed. The X-ray luminosity is $L_{\rm X}=f_{\rm X} L_{\rm bol}$. The cold thin disk is assumed to emit UV photons. As done in Proga et al. (2000) and Liu et al. (2013), we assume that the flux of UV photons is $F_{\rm UV}=2f_{\rm UV}\cos(\theta)L_{\rm bol}\exp(-\tau_{\rm UV})/4\pi r^2$, with $f_{\rm UV}$ being the ratio of UV luminosity to the bolometric luminosity. We set $\tau_{\rm UV}=\tau_{\rm X}$. We neglect the emissions in radio, optical and infrared bands, and we have $f_{\rm X}+f_{\rm UV}=1$. We assume $f_{\rm X}=0.05$ in this paper. The Compton temperature of the X-ray photons is assumed to be $T_{\rm X}= 8\times 10^7K$ (Sazonov et al. 2004).

\subsection{Hot accretion flow feedback}

\subsubsection{Wind}
It is very hard to directly observe wind from a hot accretion flow through blue-shifted absorption lines. The reason may be that hot accretion flow is fully ionized and no absorption line is present. There are some indirect observational evidences for the presence of wind produced by hot accretion flow (e.g., Crenshaw \& Kramemer 2012; Wang et al. 2013; Weng \& Zhang 2015; Cheung et al. 2016; Homan et al. 2016; Almeida et al. 2018; Park et al. 2019; Ma et al. 2019; Riffel et al. 2019). The detailed properties of wind can not be given by observations. However, we have much better theoretical understanding of the properties of wind from hot accretion flow due to the numerical simulations (e.g., Tchekhovskoy et al. 2011; Yuan et al. 2012, 2015; Narayan et al. 2012; Li et al. 2013; Almeida \& Nemmen 2019) and analytical works (e.g., Li \& Cao 2009; Gu et al. 2009; Cao 2011; Wu et al. 2013; Gu 2015; Nemmen \& Tchekhovskoy 2015; Yi et al. 2018; Kumar \& Gu 2018, 2019).

We set the wind properties from hot accretion flow according to the numerical simulation work of Yuan et al. (2015). In Bu \& Yang (2019b), we have introduced the details of how to set physical properties of wind. For convenience, we briefly introduced here. Simulations of viscous hot accretion flow (e.g., Yuan et al. 2015) find that wind is present outside $10r_s$. Because wind can take away mass, mass accretion rate decreases inwards (e.g., Yuan et al. 2012, 2015),
\begin{equation}
\dot M_{\rm inflow}(r)= \dot M_{\rm R_H}\left(\frac{r}{R_{\rm H}}\right)^{0.5}, {\rm for } \ 10r_s<r<R_{\rm H}
\label{inflowflux}
\end{equation}
with $R_{\rm H}$ being the outer boundary of the viscous hot accretion flow; $\dot M_{\rm R_H}$ is the mass inflow rate at $R_{\rm H}$. Bu et al. (2013) studied low angular momentum hot accretion flow. In Bu et al. (2013), we find that when gas from large radii has very small angular momentum, the gas will free fall from large radii to the circularization radius ($r_c$). In the region outside $r_c$, no outflows can be formed (Bu et al. 2013). When gas arrives at $r_c$, a viscous hot accretion flow will form and winds can be generated by the viscous hot accretion flow. Therefore, in the present paper, we set $R_H = r_c$ and $\dot M_{\rm {R_H}}=\dot M_{\rm in}$.
The accretion rate at black hole horizon is,
\begin{equation}
\dot M_{\rm BH}=\dot M_{\rm {R_H}}\left(\frac{10r_s}{R_{\rm H}}\right)^{0.5}
\label{inflowBH}
\end{equation}
The mass flux of wind is,
\begin{equation}
\dot M_{\rm wind}=\dot M_{\rm in}-\dot M_{\rm BH}
\label{windflux}
\end{equation}

The inner boundary of the simulation ($1500r_s$) is very close to $r_c$. Therefore, even we set $R_H =1500r_s$, according to equations (14) and (15), the mass flux of wind can not differ much from that in the case $R_H = r_c$. We have done calculations and find that if $R_H=1500r_s$, the wind mass flux will be larger than that used in the present paper by a very small factor of 0.02. Therefore, we expect that the results will not be differ much if we set $R_H = 1500r_s$. 

Comparing Equations (12) and (14), it is clear that there is a gap in the accretion rate (or luminosity) of the AGN. The maximum accretion rate for hot accretion flow is $\sim 0.2\% L_{\rm Edd}/0.1 {\rm c^2}$. According to equation (22), when accretion rate is $\sim 0.2\% L_{\rm Edd}/0.1 {\rm c^2}$, the radiative efficiency is $\sim 0.05$. Correspondingly, the maximum luminosity of hot accretion flow is $0.1\%L_{\rm Edd}$. The gap is $0.1\%-2\% L_{\rm Edd}$.

Observationally, there should be no gap in the luminosity of AGNs. The reason may be as follows. When the AGN is in the low luminosity phase, a thin disk is truncated at a radius $Rtr$. Inside $Rtr$, hot accretion flow operates. With the increase of accretion rate, $Rtr$ decreases (Liu et al. 1999; Gu \& Lu 2000; Yuan \& Narayan 2014). The accretion rate onto the black hole is equal to $\dot M_{\rm in}(10r_s/Rtr)^{0.5}$. With the decrease of $Rtr$, the accretion rate of the black hole will gradually reach the value of $\dot M_{\rm in}$. Therefore, there should be no gap in reality. In this paper, we do not consider the details of this point. However, we note that according to Figures (1), (4) and (5), the AGN spends most of their time at the luminous phase (with luminosity higher than $2\%L_{\rm Edd}$). There are only several snapshots at which AGN is in hot accretion flow mode. Therefore, the gap in luminosity of the AGN can not affect the results much.

The radial velocity of wind depends on the location where it is produced (Yuan et al. 2015),
\begin{equation}
 v_{\rm wind,r}=0.21v_k(R_{\rm wind})
\label{vwindhot}
\end{equation}
$v_k(R_{\rm wind})$ is the Keplerian velocity at wind launching radius ($R_{\rm wind}$).
According to Equation (\ref{windflux}), we have,
\begin{equation}
d \dot M_{\rm wind}/dr= d \dot M_{\rm inflow}/dr
\label{dwindflux}
\end{equation}
By using Equations (\ref{inflowflux}), (\ref{vwindhot}) and (\ref{dwindflux}), we can derive the mass flux weighted radial velocity of wind.
%As mentioned above, wind can be generated by hot accretion flow in the region $10r_s<r<R_{\rm H}$. We can calculate the mass flux weighted radial velocity of wind as follows:
%\begin{equation}
%\overline{v}_{\rm wind,r}=\frac{\int_{10r_s}^{\rm R_H} \frac {d \left(\dot M_{\rm {wind}} (r) \right)}{dr} %v_{\rm wind,r} dr}{\int_{10r_s}^{\rm R_H} \frac {d \left(\dot M_{\rm {wind}}(r) \right)}{dr}dr}
%\end{equation}
Given that $R_H=r_c=900r_s$, the mass flux weighted wind radial velocity is,
\begin{equation}
\overline{v}_{\rm wind}=0.53v_k(900r_s)
\end{equation}
with $v_k(900r_s)$ being the Keplerian velocity at $900r_s$.
$v_\theta$ of wind is found to be much smaller than both radial and rotational velocities. Therefore, we set $v_\theta$ of wind to be zero. Wind rotational velocity is also given by Yuan et al. (2015),
\begin{equation}
V_{\rm \phi, wind}=0.9v_k(R_{\rm H})
\end{equation}
Wind internal energy is found to be about 0.6 times the gravitational energy (Yuan et al. 2015),
\begin{equation}
e_{\rm wind}=\frac{0.6GM_{\rm BH}}{R_{\rm H}}
\end{equation}

The final issue is the opening angle of wind. Wind is found to be distributed in the regions $\theta \sim 30^\circ-70^\circ$ and $\theta \sim 110^\circ-150^\circ$ (Yuan et al. 2015). In the present work, only the region above midplane is calculated. Hence, wind is injected into the computational domain within $30^\circ <\theta <70^\circ$. We assume that density of wind is independent of $\theta$.

\subsubsection{Radiation}
For a LLAGN ($\dot M_{\rm in} < \dot M_c$), hot accretion flow is operating. We assume that the hot accretion flow only emits X-ray photons. In this case, $f_{\rm X}=1$ and $L_{\rm X} = L_{\rm bol}$. We also assume that the X-ray photons are spherically emitted.
Xie \& Yuan (2012) find that the radiative efficiency ($\epsilon_{\rm hot}$) depends on black hole accretion rate and the parameter $\delta$ which describes the fraction of the direct viscous heating to electrons. The radiative efficiency is as follows (Xie \& Yuan 2012):
\begin{equation}
\epsilon_{\rm hot}(\dot{M}_{\text{BH}})=\epsilon_0(\frac{100\dot{M}_{\text{BH}}}{\dot{M}_{\text{Edd}}})^a,
\end{equation}
$\dot{M}_{\text{Edd}}=L_{\rm Edd}/0.1c^2$ is the Eddington accretion rate; $\epsilon_0$ and $a$ for the case of $\delta=0.5$ can be described as:
\begin{equation}
(\epsilon_0,a) = \left\{ \begin{array}{ll}
(1.58,0.65) & \textrm{if } \frac{\dot{M}_{\text{BH}}}{\dot{M}_{\text{Edd}}}\lesssim2.9\times10^{-5};\\
(0.055,0.076) & \textrm{if } 2.9\times10^{-5}<\frac{\dot{M}_{\text{BH}}}{\dot{M}_{\text{Edd}}}\lesssim3.3\times10^{-3};\\
(0.17,1.12) & \textrm{if } 3.3\times10^{-3}<
\frac{\dot{M}_{\text{BH}}}{\dot{M}_{\text{Edd}}}\lesssim5.3\times10^{-3}.
\end{array} \right.
\end{equation}
When $\frac{\dot{M}_{\text{BH}}}{\dot{M}_{\text{Edd}}}>5.3\times10^{-3}$,
the radiative efficiency is simply set to be 0.1.
The bolometric luminosity is $L_{\rm bol}=\epsilon_{\rm hot} \dot M_{\rm BH} c^2$. The flux of X-ray photons is $F_{\rm X}=L_{\rm bol}\exp(-\tau_X)/4\pi r^2$.

The Compton temperature ($T_{\rm X}$) of the X-ray photons from hot accretion flow is calculated by Xie \& Yuan (2017). It is found $T_{\rm X} \sim 10^8$K. We set $T_{\rm X}=10^8$K, when the central engine is a LLAGN.

\subsection{Initial and boundary conditions}
Initially, we put gas with uniform density ($\rho_0$) and temperature ($T_0$) in the whole computational domain. Our resolution is $140\times 80$. In $r$ direction, in order to well resolve the inner region, we employ logarithmically spaced grid.  Grids in $\theta$ direction are uniformly spaced.

In the models with wind feedback, at the inner radial boundary, we inject wind into the computational domain. The injection $\theta$ angle and properties of the injected wind are introduced above. For the $\theta$ angle at which no wind is injected, we use outflow boundary conditions. Outflow boundary conditions do not permit gas flows into the computational domain. The physical variables in the ghost zones are just copied from the first active zone. In the models without wind feedback, at the inner radial boundary, in the whole $\theta$ angle ($0^\circ \leq \theta \leq 90^\circ$), we use outflow boundary conditions.

\begin{table*} \caption{Models and results }
\setlength{\tabcolsep}{4mm}{
\begin{tabular}{ccccccc}
\hline \hline
 Model & wind feedback & $\rho_0$ & $T_0$  & $L/L_{\rm Edd}$   & $\dot M(r_{\rm out})/\dot M_{\rm Edd}$ & $P_{\rm K}(r_{\rm out})$ \\

  &   &  ($10^{-19}\text{g cm}^{-3}$) & $ 10^6 {\rm K}$ &          &  &($L_{\rm Edd}$)   \\
(1) & (2)             & (3)                         &  (4)      &     (5)          &        (6)   & (7)   \\

\hline\noalign{\smallskip}
D19T6rad     & OFF & 1 & 2 &  28.5 & $190 $ & $0.1$  \\
D19T6windrad & ON  & 1 & 2 &  21.6 & $75$   & $9.7$  \\
D19T7rad     & OFF & 1 & 20&  21.8 & 280 & 0.07 \\
D19T7windrad & ON  & 1 & 20&  18   & 330 & 2.8 \\
D20T6rad     & OFF & 0.1 & 2&  8.7  & 5.4 & 0.02 \\
D20T6windrad & ON  & 0.1 & 2&  5.3  & 6.9 & 2.5 \\
\hline\noalign{\smallskip}
\end{tabular}}
%\end{supertabular}

Note: Col. 1: model names. Col 3: the density for initial condition. Col 4. the temperature for the initial condition. Col. 5: time-averaged luminosity (in unit of Eddington luminosity). Col. 6: time-averaged mass outflow rate at the outer boundary in unit of $\dot M_{\rm Edd}=L_{\rm Edd}/0.1\ c^2$ ($c$ is speed of light). Col. 7: time-averaged mechanical energy flux (in unit of Eddington luminosity) of outflow measured at the outer boundary. The time-average are done from 3.5 to 25 free fall timescales measured at the outer boundary.
\end{table*}

%\begin{figure*}
%\begin{center}
%\includegraphics[scale=0.5]{Luminosity.ps}\hspace*{0.7cm}
%\includegraphics[scale=0.5]{Momentum.ps}\hspace*{0.7cm}
%\hspace*{0.5cm} \caption{Left panel: powers of wind (dashed line) and radiation (solid line) generated from the central LLAGN versus black hole mass accretion rate. Power is in unit of Eddington luminosity ($L_{\rm Edd}$). Black hole mass accretion rate is in unit of Eddington accretion rate ($\dot M_{\rm Edd}=10L_{\rm Edd}/c^2$). Right panel: momentum fluxes of wind (dashed line) and radiation (solid line) generated from the central LLAGN versus black hole mass accretion rate. The momentum flux is in unit of $L_{\rm Edd}/c$. \label{Fig:power}}
%\end{center}
%\end{figure*}
In order to continuously supply gas at the outer boundary, we set the outer radial boundary conditions as follows. When we find that gas falls inward at the last active zone ($v_r<0$) at a given $\theta$ angle, then at this angle, we inject gas into the computational domain. Density, temperature, specific angular momentum of the injected gas are equal to those for the initial condition. If we find $v_r>0$ at a given $\theta$ angle, then at this angle, we employ outflow boundary conditions.

At $\theta=0$, we use axis-of-symmetry boundary conditions. At $\theta=\pi/2$, reflecting boundary conditions are employed.

\section{Results}
All of the models in this paper are summarized in Table 1.
In order to discriminate the outflow from the central AGN (not resolved in this paper) and outflow found in the simulation domain, we use ``wind" to represent the mass outflow from the central AGN, which is injected into the computational domain at inner boundary in simulations with wind feedback.  ``Outflow" is used to denote the outward moving gas found in our simulation domain.

\subsection{Effects of wind feedback}
We take the two models (D19T6rad and D19T6windrad) with $\rho_0=10^{-19}{\rm g/cm^3}$ and $T_0=2\times 10^6{\rm K}$ as our fiducial models.

\begin{figure}
\begin{center}
\includegraphics[scale=0.5]{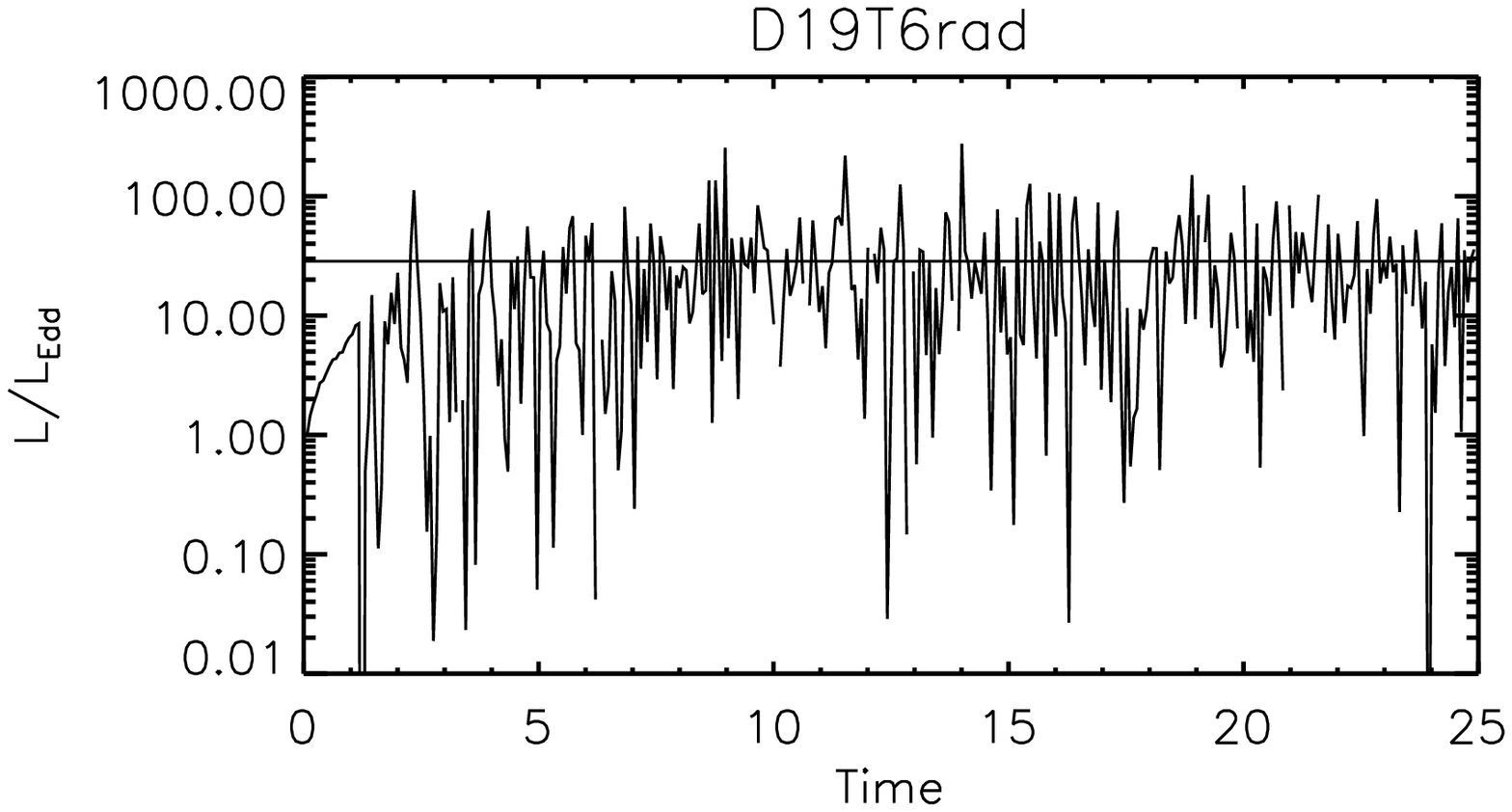}\hspace*{0.7cm} \\
\includegraphics[scale=0.5]{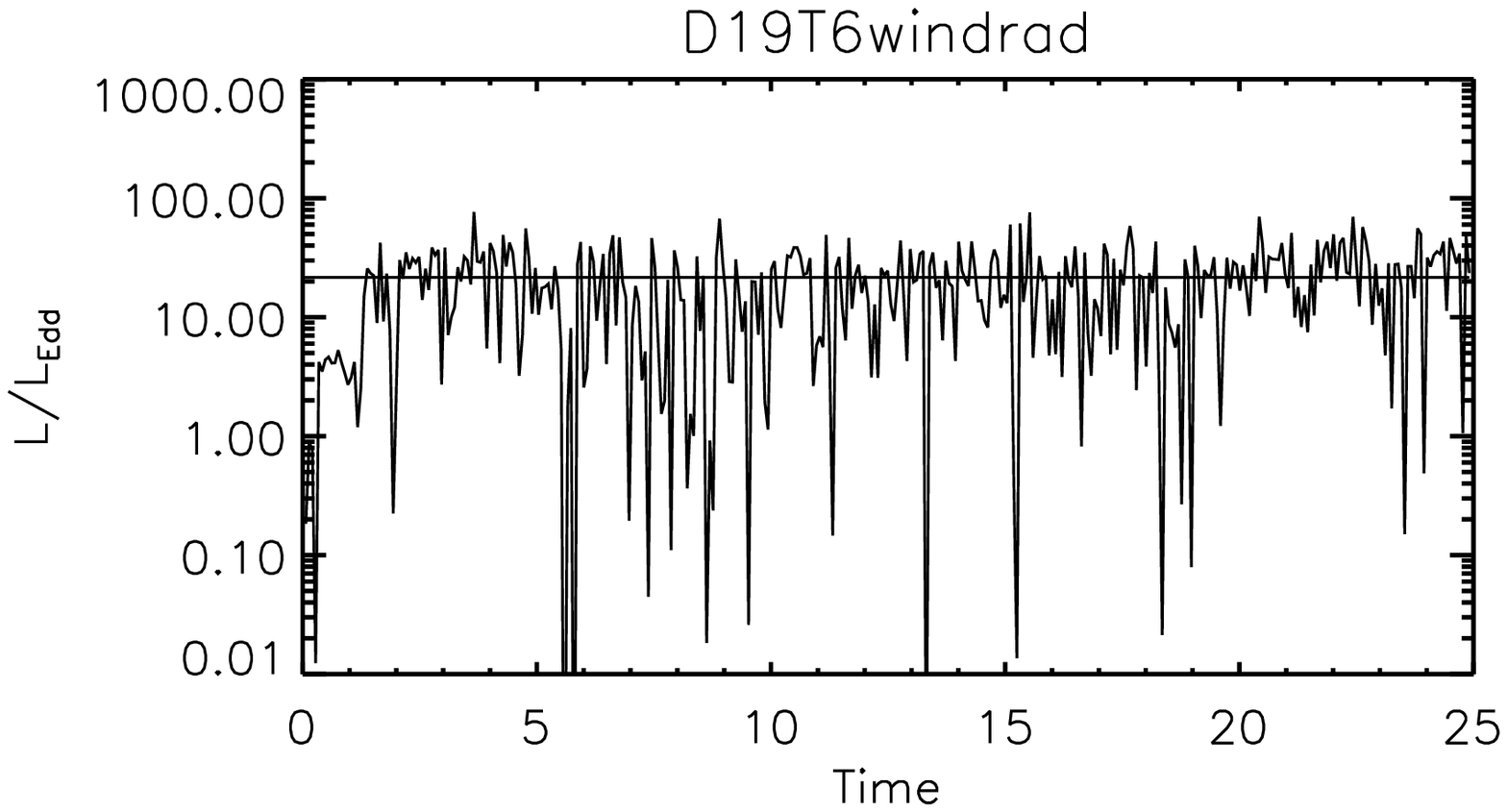}\hspace*{0.7cm}
\hspace*{0.5cm} \caption{Time evolution of the luminosity of the central AGN normalized by Eddington luminosity for models D19T6rad (top-panel) and D19T6windrad (bottom-panel). The horizontal axis is time in unit of free-fall timescale measured at the outer boundary. The horizontal solid line in each panel corresponds to the time-averaged (from $t=3.5$ to $25$ free-fall timescales measured at the outer boundary) value.  \label{Fig:L19T6}}
\end{center}
\end{figure}

\begin{figure*}
\begin{center}
\includegraphics[scale=0.5]{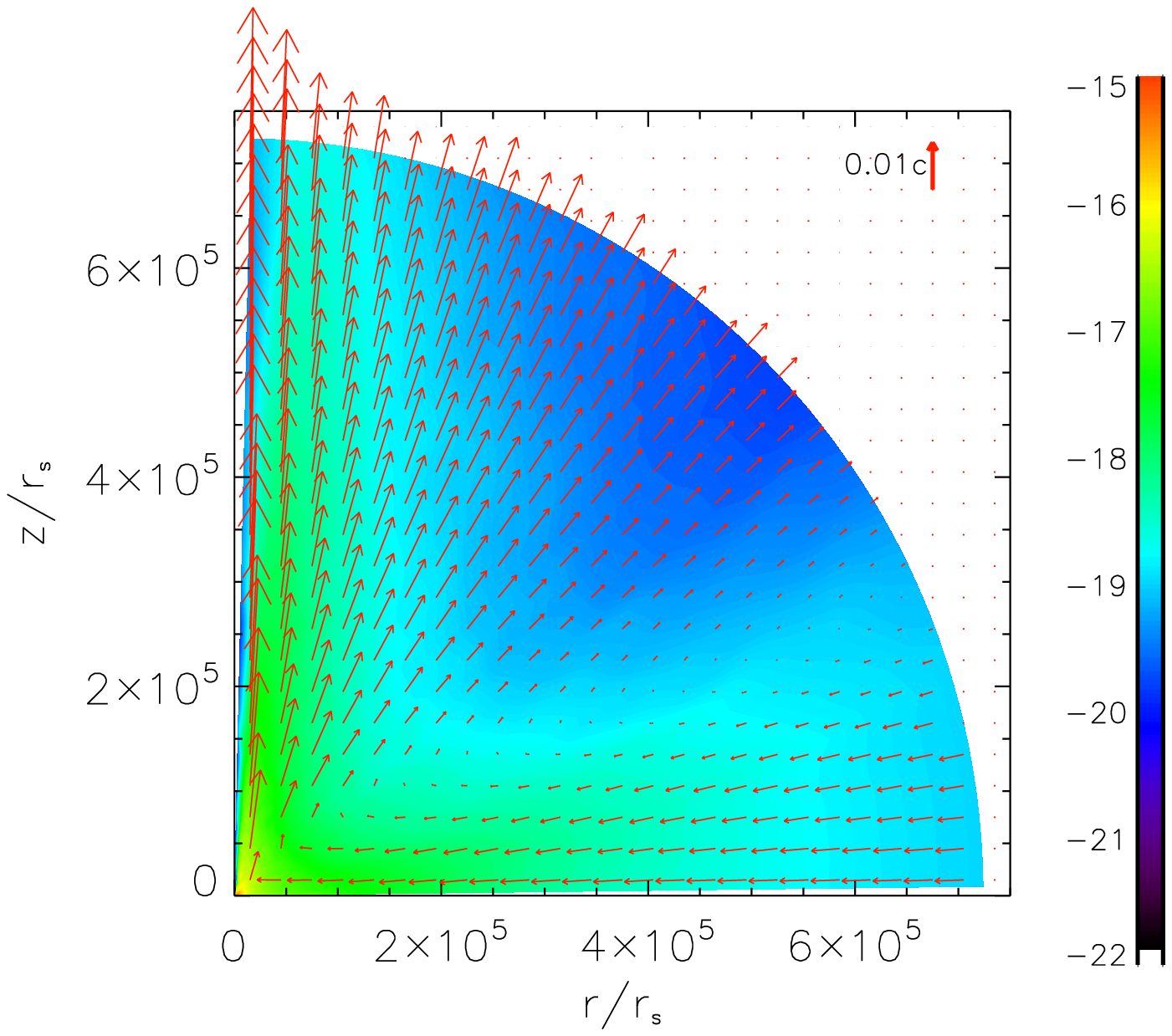}\hspace*{0.1cm}
\includegraphics[scale=0.5]{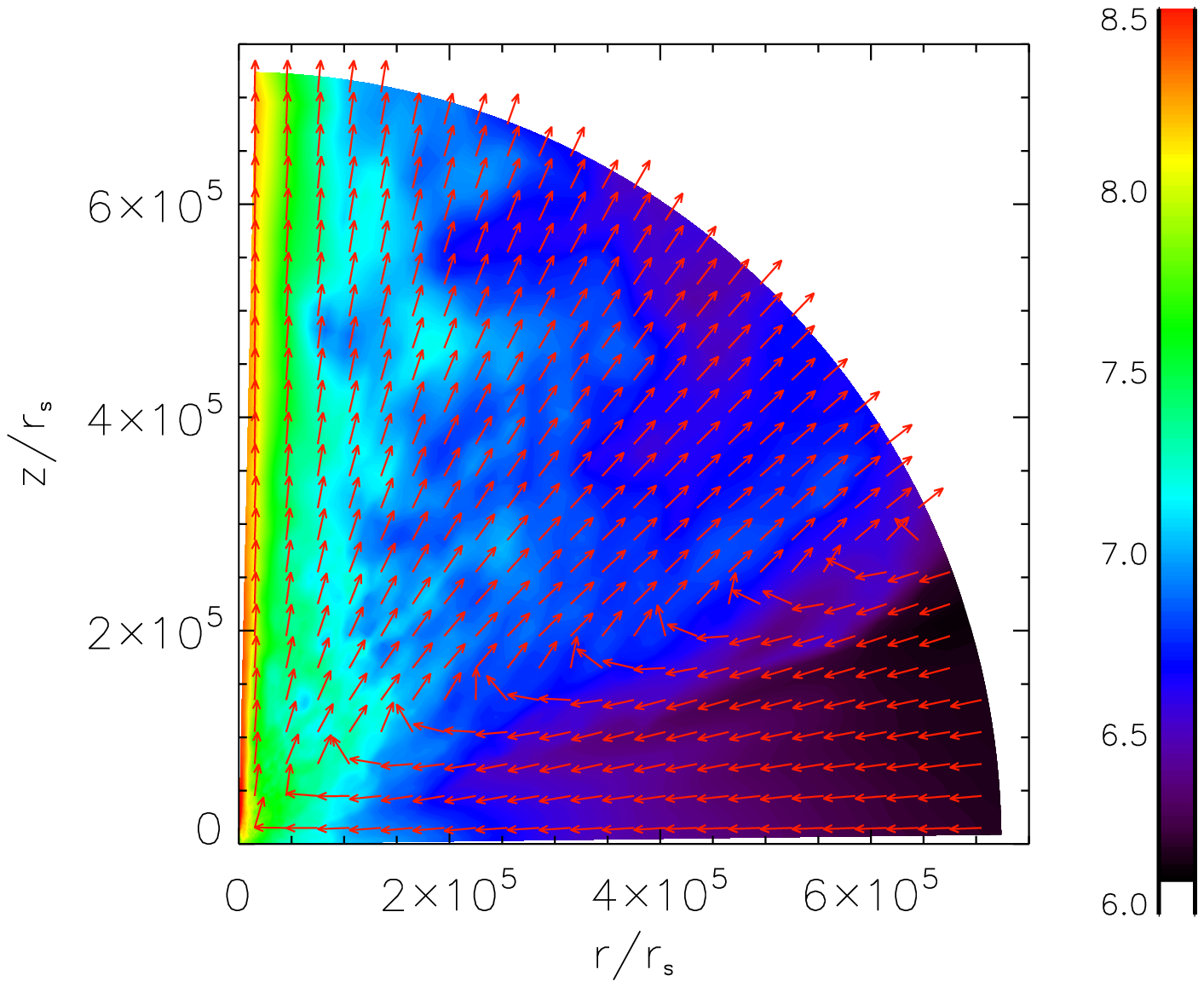}\hspace*{0.7cm}\\
\includegraphics[scale=0.5]{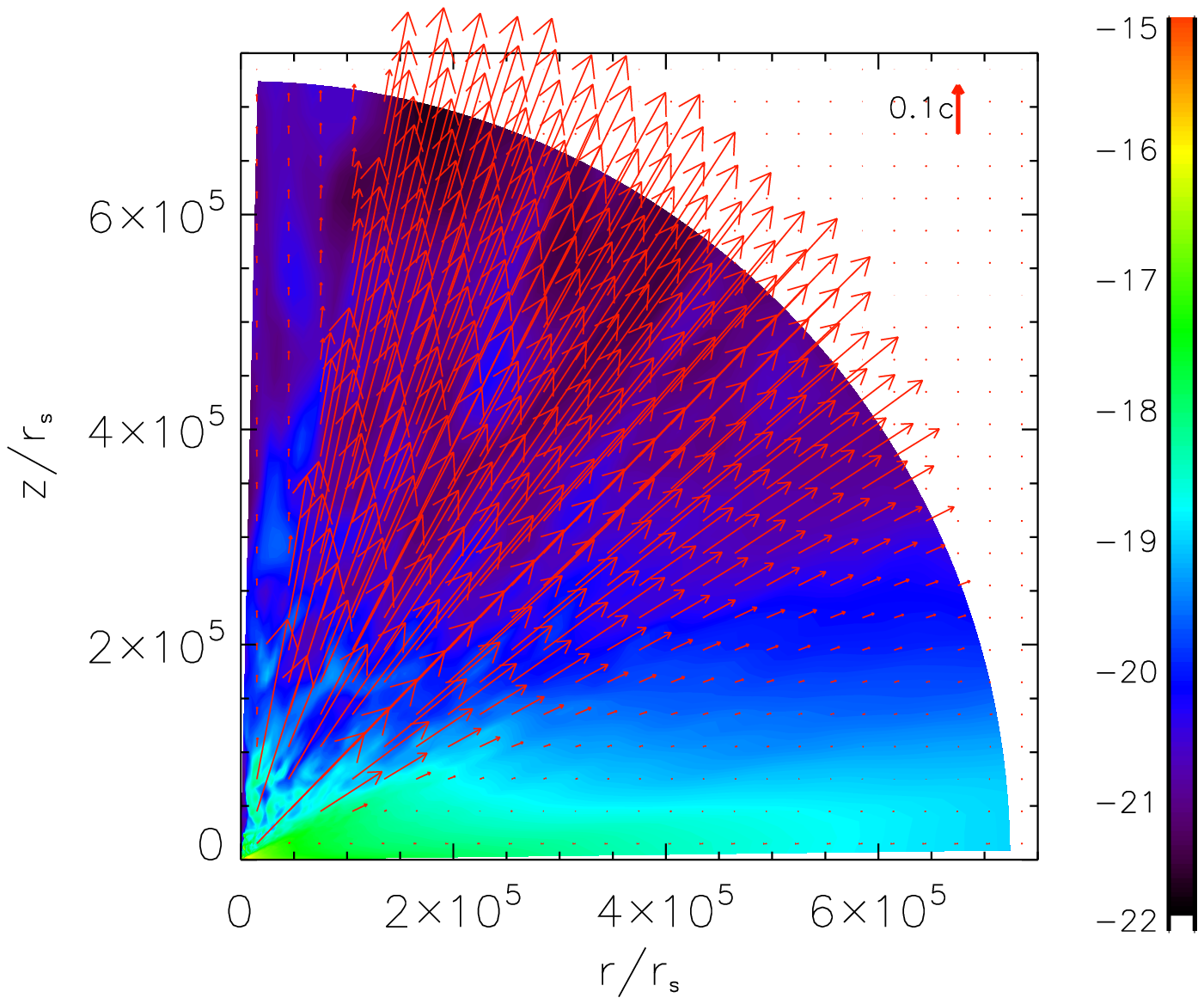}\hspace*{0.1cm}
\includegraphics[scale=0.5]{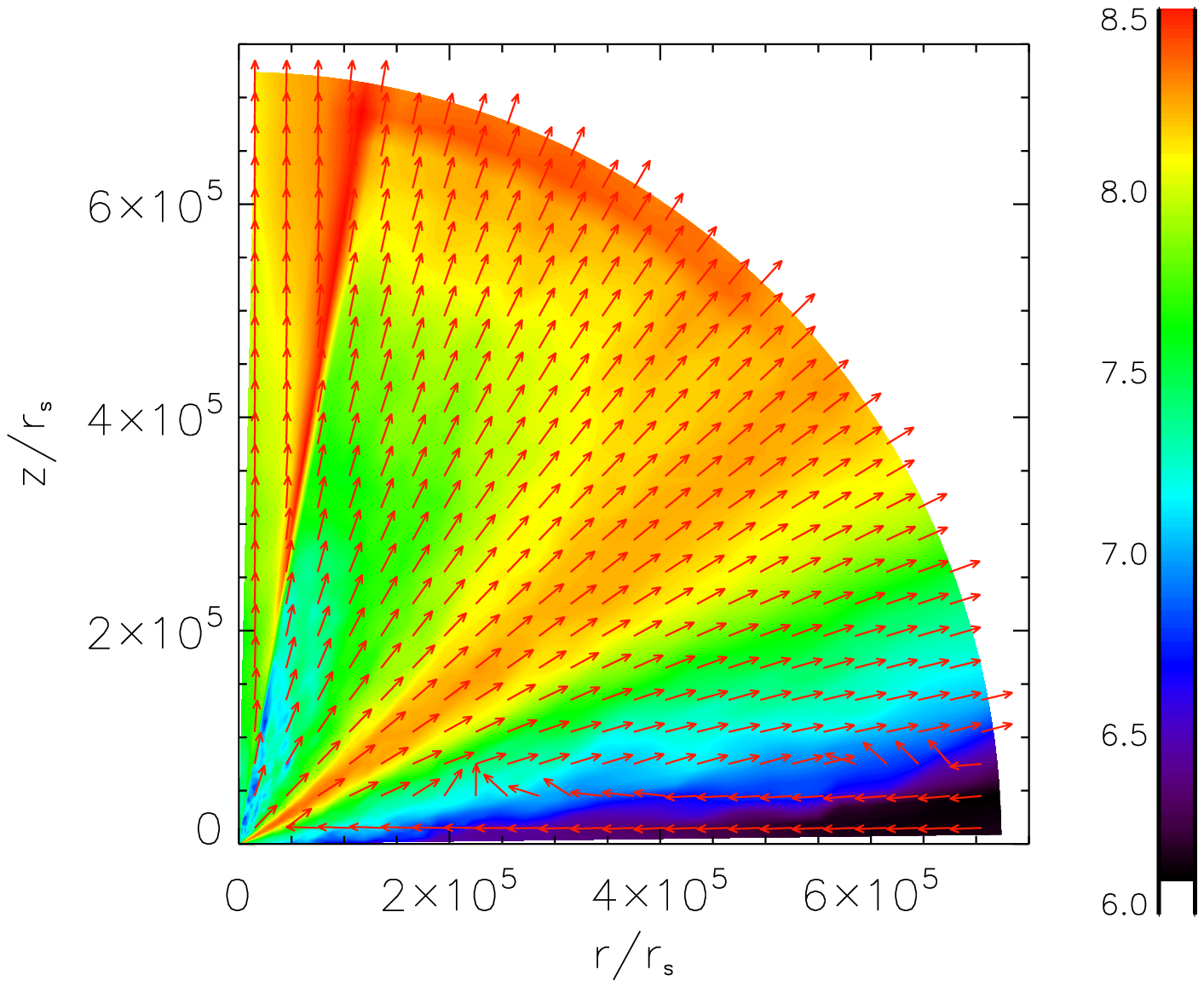}\hspace*{-0.4cm}
\hspace*{0.5cm} \caption{Two-dimensional time-averaged (from $t=3.5$ to $25$ free fall timescale measured at outer boundary) properties for models D19T6rad (top panels) and D19T6windrad (bottom panels). The left column panels plot logarithm density (color) and velocity vector (arrows). The right column panels plot logarithm temperature (color) and unit vector of velocity (arrows). \label{Fig:vector19}}
\end{center}
\end{figure*}

Figure \ref{Fig:L19T6} shows the time evolution of luminosity of the central AGN in unit of $L_{\rm Edd}$. The top and bottom panels are for models D19T6rad and D19T6windrad, respectively. The horizontal solid line in each panel is the time-averaged value. The luminosity in both models fluctuates significantly. The highest luminosity in model D19T6rad is $\sim 300L_{\rm Edd}$. The lowest luminosity can be lower than $1\%L_{\rm Edd}$. In model D19T6rad, the fluctuation is due to the radiative feedback (radiation pressure) from the central AGN. The reason for the fluctuation is as follows. When the luminosity is significantly higher than $L_{\rm Edd}$,  radiation pressure due to Thomson scattering can be significantly exceeding gravity. Outflows can be launched and gas around the central AGN is taken away. Then the accretion rate (or luminosity) decreases and radiative feedback becomes weak. The gas around the AGN will fall back and the AGN will be triggered again. The time-averaged luminosity for model D19T6rad is $28.5 L_{\rm Edd}$.

The luminosity evolution patten in model D19T6windrad is very similar as that in model D19T6rad. The fluctuation of luminosity in model D19T6windrad is due to both wind and radiative (pressure) feedback. The wind feedback does decrease the luminosity of the central AGN. The luminosity of the central AGN in model D19T6windrad has never being higher than $100L_{\rm Edd}$. The time-averaged luminosity in model D19T6windrad is $21.6 L_{\rm Edd}$. After taking into account wind feedback, the decrease of luminosity is not significant.

\begin{figure}
\begin{center}
\includegraphics[scale=0.5]{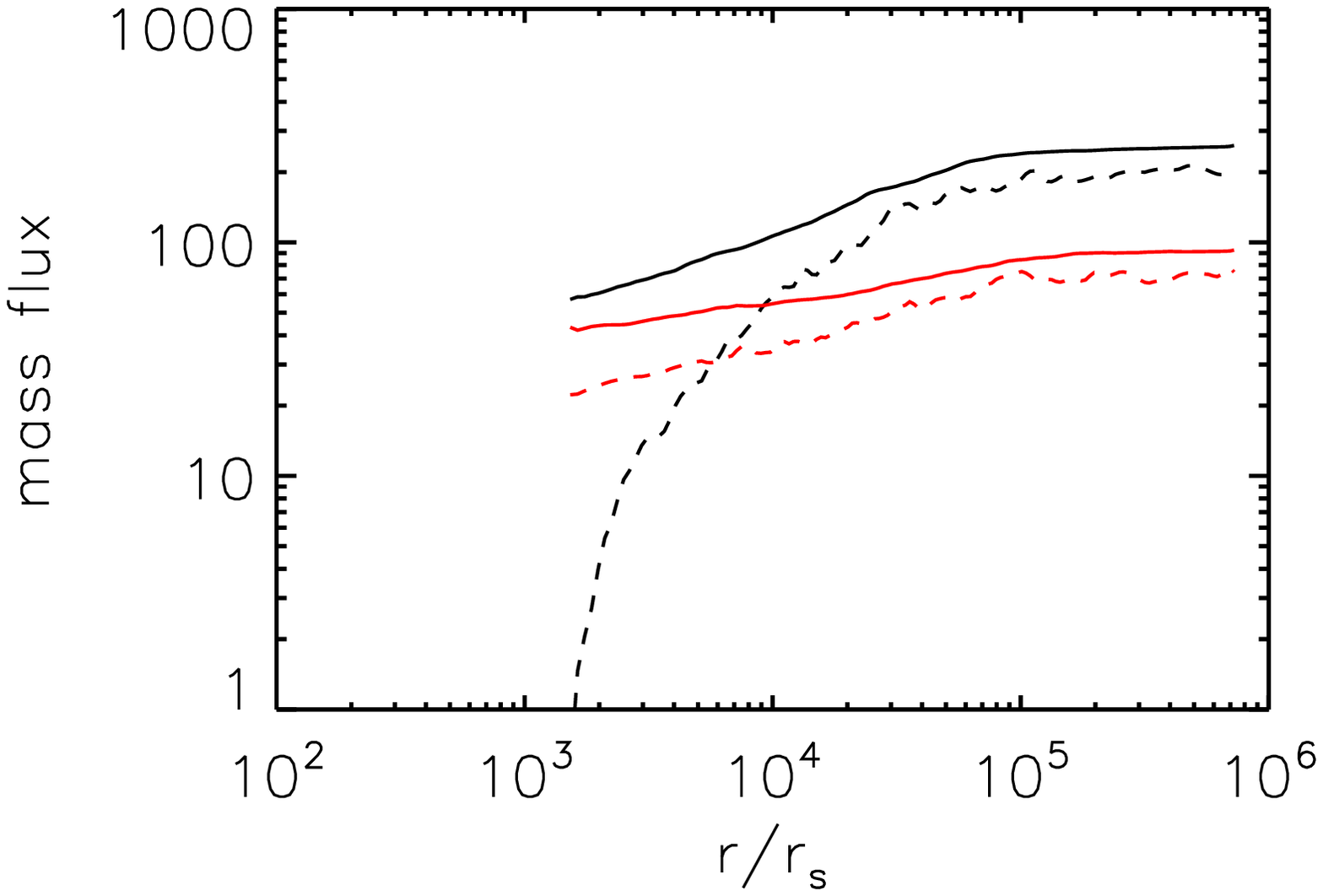}\hspace*{0.7cm}
\hspace*{0.5cm} \caption{Radial profiles of time-averaged (from 3.5 to $25$ free-fall timescales measured at the outer boundary) mass inflow rate (solid lines) and outflow rate (dashed lines) for models D19T6rad (black lines) and D19T6windrad (red lines). The mass fluxes are in unit of Eddington accretion rate ($\dot M_{\rm Edd}=L_{\rm Edd}/0.1c^2$). \label{Fig:massflux19}}
\end{center}
\end{figure}

\begin{figure}
\begin{center}
\includegraphics[scale=0.5]{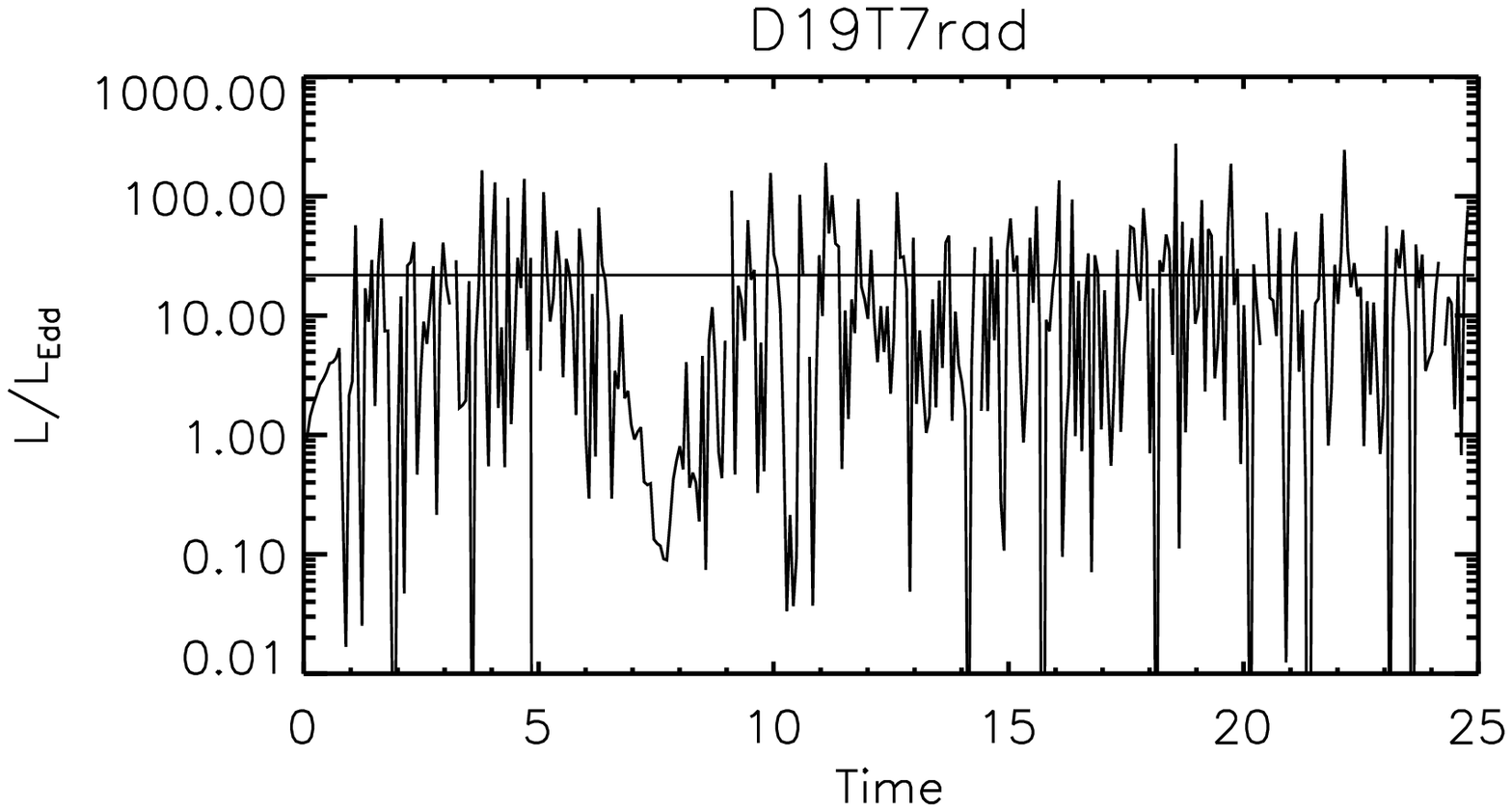}\hspace*{0.7cm} \\
\includegraphics[scale=0.5]{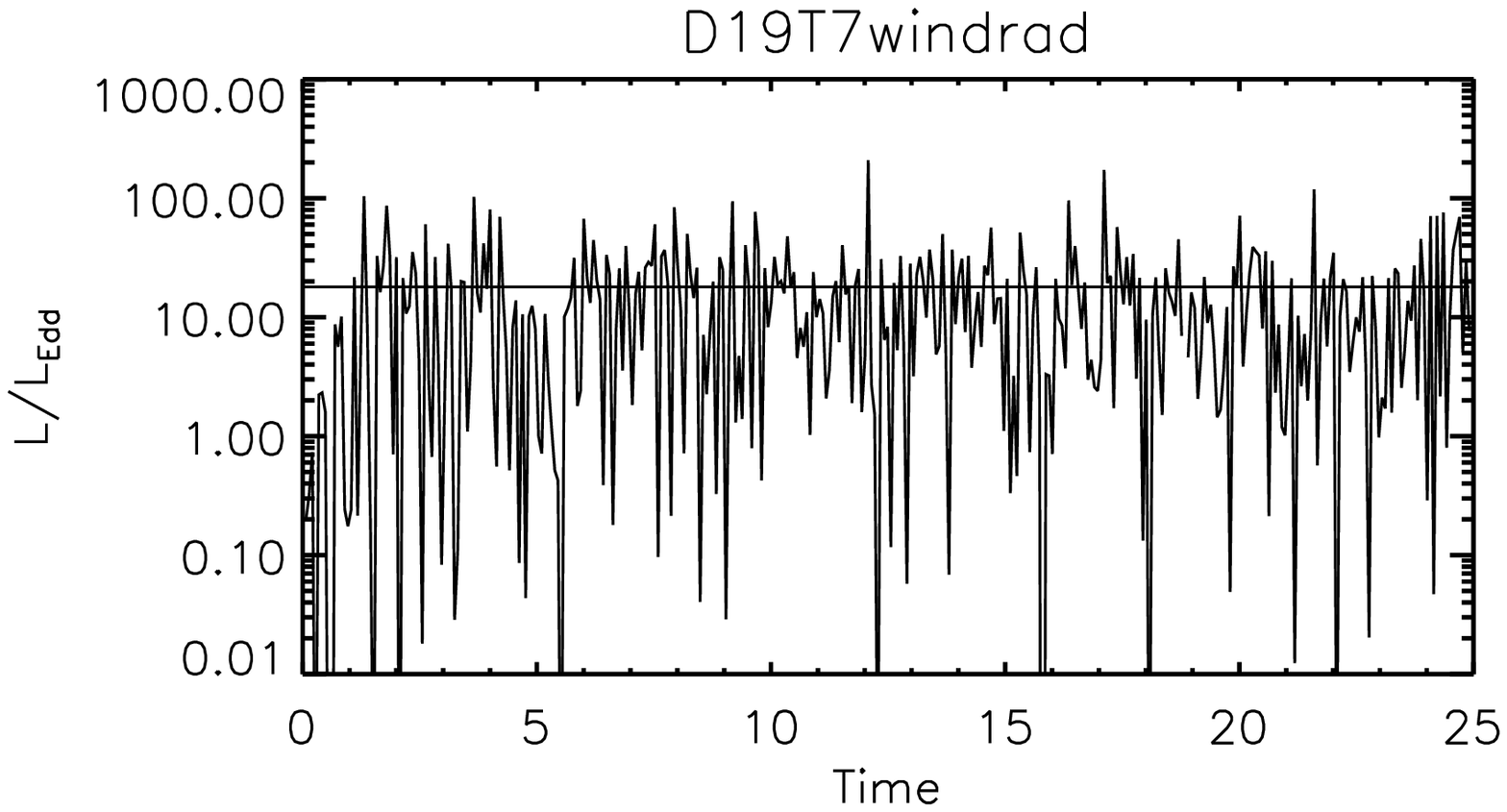}\hspace*{0.7cm}
\hspace*{0.5cm} \caption{Same as Figure \ref{Fig:L19T6}, but for models D19T7rad (top-panel) and D19T7windrad (bottom-panel). \label{Fig:L19T7}}
\end{center}
\end{figure}

In order to study the detailed properties of the flow, we plot Figure \ref{Fig:vector19}. In this figure, we show the time-averaged (from 3.5 to 25 free-fall timescales measured at the outer boundary) two-dimensional structure for models D19T6rad (top-panels) and D19T6windrad (bottom panels). The left column panels plot logarithm density (color) and velocity vector (arrows). The right column panels plot logarithm temperature (color) and unit vector of velocity (arrows). In model D19T6rad, inflow exists in the region around midplane ($65^\circ < \theta \leq 90^\circ$). In the region $\theta < 65^\circ$, outflow is present. The outflow velocity increases with decreasing of $\theta$. In this model, outflows are launched due to radiation (mainly UV photons) pressure due to Thomson scattering. Because the UV photon flux from the central AGN has a $\sin\theta$ dependence (see Section 2.2.2), the radiation force per unit mass increases with decrease of $\theta$. Therefore, the outflow velocity increases with decrease of $\theta$. The maximum velocity of outflow is $\sim 0.02 c$.

In model D19T6windrad, we take into account wind feedback. From the bottom panel of Figure \ref{Fig:L19T6}, we see that the central AGN spends most of its time in luminous AGN phase. For the luminous AGN, wind is injected into the computational domain in the region $10^\circ < \theta < 70^\circ$ (see Section 2.2.1). In addition to radiation pressure, outflows can also be driven by wind feedback. There are several differences between the properties of outflows in models D19T6rad and D19T6windrad. First, in model D19T6windrad, the outflows have much larger opening angle ($0^\circ < \theta < 82^\circ$). Second, the density in the outflow region in model D19T6windrad is much lower. The reason is that wind from the central AGN can directly blow away gas, which results in a much lower density in the outflow region. Third, the gas temperature in the outflow region in model D19T6windrad is significantly higher than that in the outflow region in model D19T6rad. The are two reasons. The first one is that some portion of the energy of the wind from the central AGN is transferred to the gas and becoming gas internal energy. The gas in the outflow region is heated by the compression work done by wind. The second one is that lower gas density in outflow region in model D19T6windrad results in lower radiative cooling rate. Finally, the maximum velocity of outflow in model D19T6windrad is $\sim 0.2-0.3c$, which is one order of magnitude higher than that of outflow in model D19T6rad.

In Figure \ref{Fig:massflux19}, we plot the radial profiles of time-averaged (from 3.5 to 25 free-fall timescales measured at outer boundary) mass inflow (solid lines) and outflow (dashed lines) rates for models D19T6rad (black lines) and D19T6windrad (red lines). In both of the models, due to the presence of outflow, the mass inflow rate decreases inwards. The mass outflow rate in both models increases with increasing of radius. This means that when outflow moves outwards, more and more inflow gas joins outflows. The mass inflow rate in model D19T6rad is larger. The reasons are as follows. First, in model D19T6rad, the inflow occupies larger solid angle at any radius (see Figure \ref{Fig:vector19}). Second, the density of the inflow gas in model D19T6rad is higher (see Figure \ref{Fig:vector19}). Finally, the infall velocity of inflowing gas in the two models is comparable. In the region $r>6000r_s$, the mass outflow rate in model D19T6windrad is lower. The reason is as follows. Even though the velocity of outflow is higher in model D19T6windrad, the significant lower outflow density (see Figure \ref{Fig:vector19}) makes the mass outflow rate in this model much lower.

From Figure \ref{Fig:massflux19}, we can see that at the inner boundary, the mass inflow rate in model D19T6rad does not differ much from that in model D19T6windrad. The accretion rate onto the black hole is half of the mass inflow rate at the inner boundary (see Section 2.2.1). The mass accretion rates onto the black hole in these two models do not differ much. Consequently, the luminosity of the central AGN does not change much by wind feedback (see Figure \ref{Fig:L19T6}). Liu et al. (2013) studied the accretion flow with exactly same initial and boundary conditions as those in models D19T6rad and D19T6windrad. In Liu et al. (2013), wind feedback has not been taken into account. They find that the radiative feedback (including radiation pressure due to Thomson scattering and line force) can not make the luminosity of the central AGN sub-Eddington. The results in the present paper demonstrate that the presence of wind and radiative (without line force) feedbacks can also not solve the `sub-Eddington' puzzle. Therefore, in order to make the AGN's luminosity sub-Eddington, we may simultaneously consider wind feedback, radiation pressure due to Thomson scattering and line force. In addition, in this paper, we do not consider dust. In the region at parsec scale, dust may exist (Wang 2008; Wang et al. 2017; Czerny et al. 2016, 2019). Dust can absorb the radiation from the central AGN and then emits infrared photons. The emitted infrared photons can exert a strong radiation pressure on dust (Dorodnitsyn \& Kallman 2012; Dorodnitsyn et al. 2016). This may also help produce strong wind and decrease the luminosity of the central AGN. In future, it is very necessary to perform simulations by taking into account wind feedback, radiation pressure due to Tomson scattering and line force, and radiation pressure on dust simultaneously to study this problem again.

\begin{figure}
\begin{center}
\includegraphics[scale=0.5]{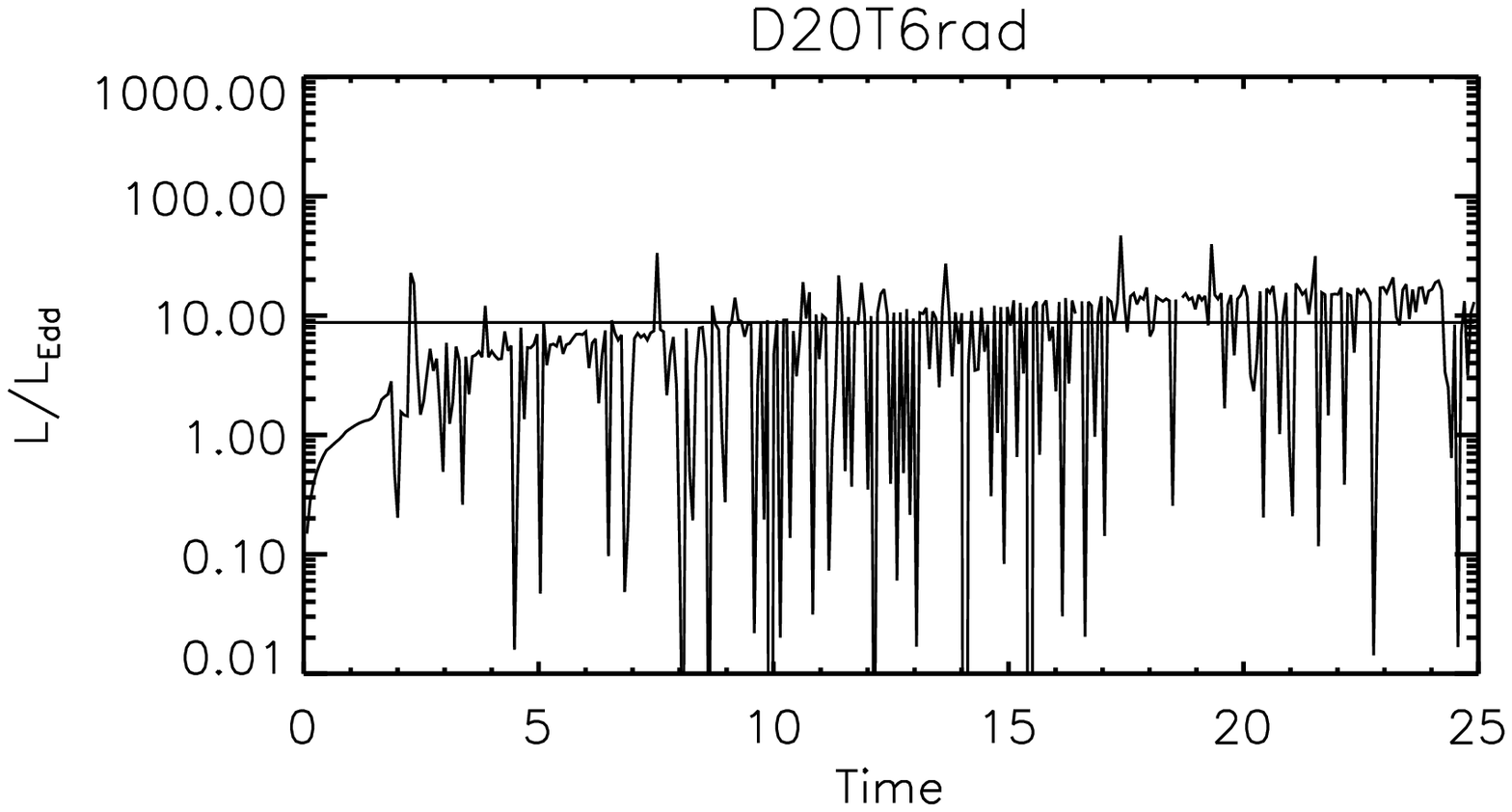}\hspace*{0.7cm} \\
\includegraphics[scale=0.5]{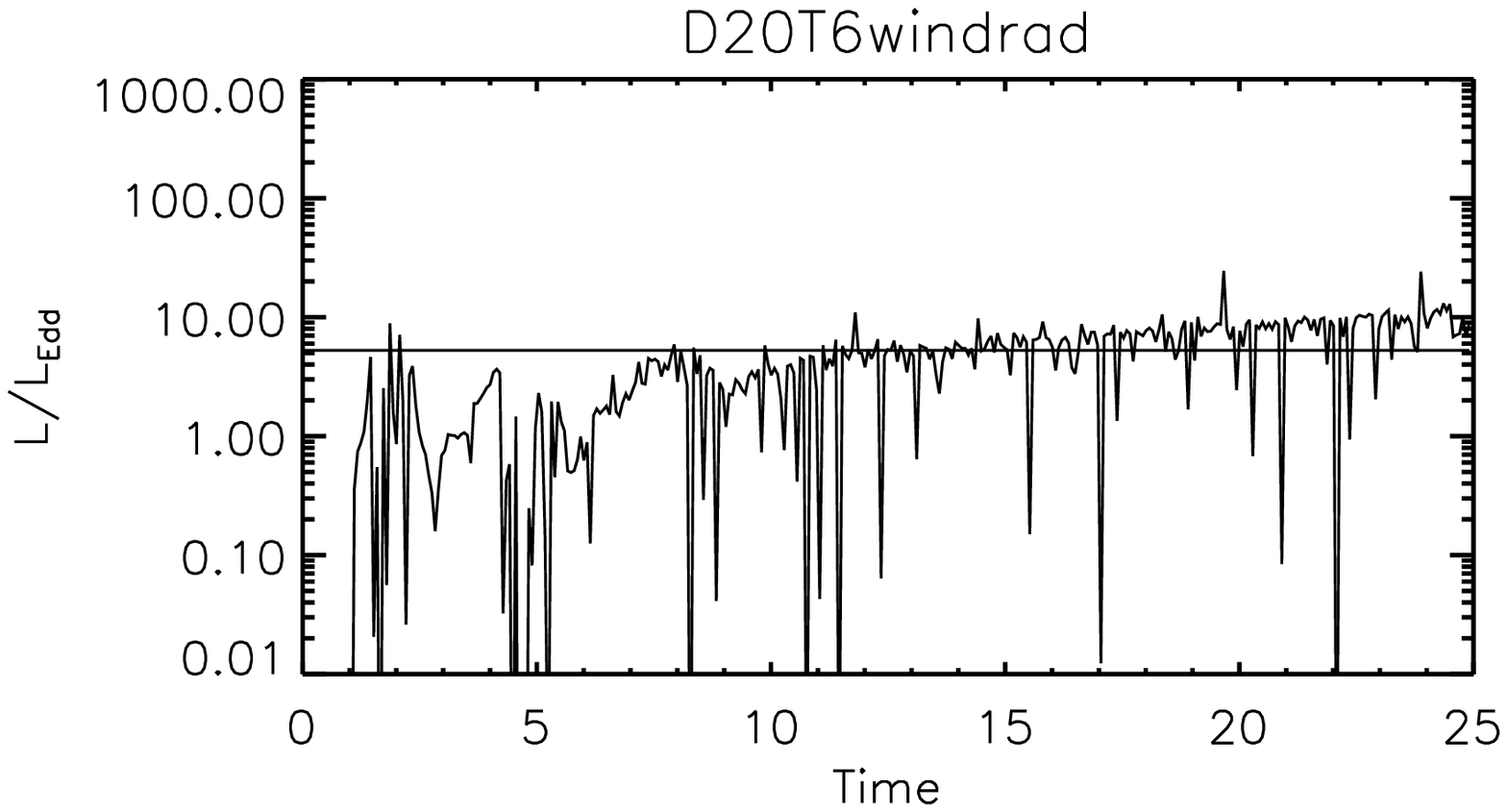}\hspace*{0.7cm}
\hspace*{0.5cm} \caption{Same as Figure \ref{Fig:L19T6}, but for models D20T6rad (top-panel) and D20T6windrad (bottom-panel). \label{Fig:L20T6}}
\end{center}
\end{figure}

One of the most important properties of outflows in AGN feedback study is the kinetic power. The kinetic power of outflow is as follows,
\begin{equation}
P_{\rm K} (r)=2\pi r^2 \int_{\rm 0^\circ}^{\rm 90^\circ} \rho
\max(v_r^3,0) \sin\theta d\theta
\end{equation}
We list the kinetic power (in unit of Eddington Luminosity) of outflows calculated at outer boundary for all of our models in column 7 of Table 1.
The kinetic power $P_{\rm K} \propto \dot M_{\rm outflow} v_{\rm outflow}^2$, with $\dot M_{\rm outflow}$ and $v_{\rm outflow}$ being mass flux and velocity of outflow, respectively. From Figure \ref{Fig:massflux19}, we see at the outer boundary, the outflow mass flux in model D19T6rad is $2 \sim 3$ times higher than that in model D19T6windrad. However, the outflow velocity in model D19T6windrad is more than one order of magnitude higher than that of outflow in model D19T6rad. Therefore, the outflow kinetic power in model D19T6windrad is almost 2 orders of magnitude higher than that of outflow in model D19T6rad.

\subsection{Dependence on gas temperature at the outer boundary}
In models D19T7rad and D19T7windrad, if gas flows inwards at the outer boundary, then we set the gas temperature at outer boundary to be one order of magnitude higher than that in models D19T6rad and D19T6windrad. In Figure \ref{Fig:L19T7}, we plot the time evolution of luminosity of the central AGN. As in the models introduced above, due to the AGN feedback, the luminosity in models D19T7rad and D19T7windrad fluctuates significantly. The time-averaged luminosity of the central AGN is just slightly decreased by AGN wind feedback. As in models with other values of density and temperature, the mass flux of outflow at the outer boundary does not be affected much by wind feedback (see the sixth column of Table 1). The kinetic power of outflow measured at the outer boundary is significantly enhanced by wind feedback (see the seventh column of Table 1). We also studied the properties of outflow (density, temperature and velocity). The effects of wind feedback are also very similar as those in models with $\rho_0=10^{-19} {\rm g \ cm^{-3}}$ and $T_0=2\times 10^6 {\rm K}$. For example, compared to model without wind feedback (D19T7rad), in the model with wind feedback (D19T7windrad), density in the outflow region is significantly lower, outflow temperature is much higher, velocity of outflows is significantly higher.

\subsection{Dependence on gas density at the outer boundary}
In models D20T6rad and D20T6windrad, if gas flows inwards at the outer boundary, then we set the gas density at outer boundary to be one order of magnitude lower than that in models D19T6rad and D19T6windrad. In Figure \ref{Fig:L20T6}, we plot the time evolution of luminosity of the central AGN. As in models with other different values of density and temperature, the wind feedback has very small effects on the time-averaged luminosity of the central AGN. Also, the mass fluxes of outflows at the outer boundary do not differ much in models with and without wind feedback (see the sixth column of Table 1). The kinetic power of outflow measured at the outer boundary in model with wind feedback is roughly 2 orders of magnitude higher than that of outflow in model without wind feedback (see the seventh column of Table 1). The effects of wind feedback on properties of outflow are also similar as those in the models introduced above. In the model with wind feedback (D20T6windrad), temperature in the outflow region is much higher, density in the outflow region is much lower, the velocity of outflow is significantly higher.

\section{Summary}
Two-dimensional hydrodynamic simulations are performed to study the effects of AGN radiation and wind feedback on the properties of slowly rotating accretion flow at sub-parsec to parsec scale. We find that when only radiation feedback is considered, outflows can be launched by the radiation pressure force due to Thomson scattering. The maximum velocity of the outflow is $\sim 0.02 c$. Due to the presence of outflow, the mass inflow rate decrease inwards. The mass flux and kinetic power of outflows depend on the density and temperature of gas at parsec scale.

We find that when wind feedback is taken into account. the mass flux of outflow can not be significantly affected. Accordingly, the luminosity of the central AGN does not change much after taking into account wind feedback. The `sub-Eddington' puzzle can not be solved by considering wind feedback. However, wind feedback can affect other properties of outflow significantly. For example, maximum velocity of outflow can be increased by a factor of $\sim 100$ by the wind feedback. Consequently, the kinetic power of outflow can be increased by a factor of $40 - 100$. Other properties of outflow such as density and temperature are also significantly affected by wind feedback. In the model with wind feedback, the density of outflow is significantly lower, the temperature of outflow is significantly higher.

\section*{Acknowledgments}
D.-F. Bu is supported in part the Natural Science Foundation of China (grants 11773053,
11573051, 11633006 and 11661161012), and the Key
Research Program of Frontier Sciences of CAS (No. QYZDJSSW-
SYS008). X.-H. Yang is supported in part by the Fundamental Research Funds for the Central Universities (No.  106112015CDJXY300005 and No. 2019CDJDWL0005) and the Natural Science Foundation of China (grant 11847301).  This work made use of the High Performance Computing Resource in the Core
Facility for Advanced Research Computing at Shanghai Astronomical
Observatory.

\end{document}